\documentclass[12pt]{iopart}
\usepackage{iopams}  
\usepackage{graphicx}

\newtheorem{cor}{Corollary}
\newtheorem{theorem}{Theorem}
\newtheorem{remark}{Remark}

\begin{document}

\title[Rotating Bowen--York initial data with a positive cosmological constant]{Rotating Bowen--York initial data with a positive cosmological constant}

\author{Patryk Mach}

\address{Institute of Physics, Jagiellonian University, \L{}ojasiewicza 11, 30-438 Krak\'{o}w, Poland}
\ead{patryk.mach@uj.edu.pl}

\author{Jerzy Knopik}
\address{Gravitational Physics, University of Vienna, Boltzmanngasse 5, 1090 Vienna, Austria}
\address{Institute of Physics, Jagiellonian University, \L{}ojasiewicza 11, 30-438 Krak\'{o}w, Poland}
\ead{jerzy.knopik@univie.ac.at}

\vspace{10pt}

\begin{abstract}
A generalization of the Bowen--York initial data to the case with a positive cosmological constant is investigated. We follow the construction presented recently by Bizo\'{n}, Pletka and Simon, and solve numerically the Lichnerowicz equation on a compactified domain $\mathbb S^1 \times \mathbb S^2$. In addition to two branches of solutions depending on the polar variable on $\mathbb S^2$ that were already known, we find branches of solutions depending on two variables: the polar variable on $\mathbb S^2$ and the coordinate on $\mathbb S^1$. Using Vanderbauwhede's results concerning bifurcations from symmetric solutions, we show the existence of the corresponding bifurcation points. By linearizing the Lichnerowicz equation and solving the resulting eigenvalue problem, we collect numerical evidence suggesting the absence of additional branches of solutions.
\end{abstract}

%
%

\section{Introduction}

In a recent paper Bizo\'{n}, Pletka and Simon investigated Bowen--York type initial data for the Einstein equations with a positive cosmological constant \cite{BPS}. Bowen--York initial data are obtained within the framework of the so-called conformal method. From the technical point of view, its most important part amounts to solving a corresponding Lichnerowicz equation. In this paper we find numerically new, symmetry-breaking branches of solutions of the Lichnerowicz equation that describes the `rotating' case investigated in \cite{BPS}. Their existence was predicted in \cite{BPS}. Basing on a bifurcation theorem due to Vanderbauwhede \cite{vanderbauwhede}, we formalise the arguments presented in \cite{BPS} and show the existence of bifurcation points that give rise to these symmetry-breaking branches of solutions. We also collect numerical evidence suggesting the absence of additional solutions.

We are interested in the initial data consisting of a triple $(\tilde {\mathcal M}, \tilde g_{ij}, \tilde K_{ij})$, where $\tilde {\mathcal M}$ is a 3-dimensional compact manifold, $\tilde g_{ij}$ is a smooth Riemannian metric, and $\tilde K_{ij}$ is a trace-free ($\tilde g^{ij}\tilde K_{ij} = 0$) tensor satisfying the Einstein vacuum constraint equations
\begin{equation}
\label{einsteinid}
\tilde R - \tilde K_{ij} \tilde K^{ij} - 2 \Lambda = 0, \quad \tilde \nabla_i \tilde K^{ij} = 0.
\end{equation}
Here $\Lambda$ is a positive cosmological constant, $\tilde R$ is the scalar curvature of $(\tilde {\mathcal M}, \tilde g_{ij})$, and $\tilde \nabla_i$ is the corresponding covariant derivative with respect to the metric $\tilde g_{ij}$.

The conformal method of finding such initial data can be summarized as follows. Let $(\mathcal M, g_{ij})$ be a compact 3-dimensional Riemannian manifold with a smooth metric $g_{ij}$ in the positive Yamabe class. In addition, let $K_{ij}$ be a smooth trace-free ($g^{ij}K_{ij} = 0$) and divergence-free ($\nabla_i K^{ij} = 0$) tensor on $\mathcal M$. The initial data $(\tilde {\mathcal M}, \tilde g_{ij}, \tilde K_{ij})$ can be found as
\[ \tilde g_{ij} = \phi^4 g_{ij}, \quad \tilde K^{ij} = \phi^{-10} K^{ij},  \]
where the conformal factor $\phi$ is a positive, smooth solution of the Lichnerowicz equation
\begin{equation}
\label{lich1}
 - \Delta_g \phi + \frac{1}{8}R \phi - \frac{1}{4} \Lambda \phi^5 - \frac{\Omega^2}{8 \phi^7} = 0.
\end{equation}
In the above formulas $\Delta_g$, $\nabla_i$, and $R$ denote the Laplacian, the covariant derivative and the scalar curvature with respect to metric $g_{ij}$. Note that setting $\tilde K^{ij} = \phi^{-10} K^{ij}$ is equivalent to $\tilde K_{ij} = \phi^{-2} K_{ij}$. Following \cite{BPS} we also denote $\Omega^2 = K^{ij}K_{ij}$.

Clearly, apart from the technical difficulty of solving Eq.\ (\ref{lich1}) for the conformal factor, the key point of the above conformal method is to find an appropriate `seed' metric $g_{ij}$ together with the trace- and divergence-free tensor $K_{ij}$. A simple, but physically important choice was introduced by Bowen and York in 1980 \cite{bowen_york}, for the case with vanishing cosmological constant. Their initial data are obtained by choosing as a `seed' manifold $\mathcal M = \mathbb R^3 \setminus \{ 0 \}$ (which is noncompact) endowed with the flat Euclidean metric. The original forms of $K_{ij}$ given by Bowen and York were later generalized by Beig \cite{beig}. In this latter version they read, in Cartesian coordinates $(x^1,x^2,x^3)$,
\begin{eqnarray}
K_{ij}(x) & = & \frac{3}{2r^2} \left[ P_i n_j + P_j n_i - (\delta_{ij} - n_i n_j) P^k n_k \right], \\
K_{ij}(x) & = & \frac{3}{r^3} \left( \epsilon_{kli} J^k n^l n_j + \epsilon_{klj} J^k n^l n_i \right), \label{byrot}  \\
K_{ij}(x) & = & \frac{C}{r^3} (3 n_i n_j - \delta_{ij}), \\
K_{ij}(x) & = & \frac{3}{2r^4} \left[ -Q_i n_j - Q_j n_i - (\delta_{ij} - 5n_i n_j) Q^k n_k \right].
\end{eqnarray}
Here $r = \sqrt{x_i x^i}$, $n_i = x_i/r$; $P_i$, $S_i$ and $Q_i$ are constant; $\delta_{ij}$ and $\epsilon_{klj}$ denote the Kronecker delta and the usual 3-dimensional permutation symbol, respectively. Vectors $P_i$ and $S_i$ can be interpreted respectively as the Arnowitt--Deser--Misner (ADM) linear and angular momenta at $r \to \infty$. In this paper we are only interested in the rotational case (\ref{byrot}).

A transformation of the Bowen--York data (\ref{byrot}) to the case with a positive cosmological constant can be done along the lines of the derivation described in \cite{BPS}. The flat metric $g = dr^2 + r^2 (d\theta^2 + \sin^2 \theta d \varphi^2)$ is first transformed conformally into
\[ \hat g = \hat \phi^4 g = \hat \phi^4 \left[ dr^2 + r^2 (d\theta^2 + \sin^2 \theta d \varphi^2) \right] \]
with $\hat \phi^4 = 1/(\Lambda r^2)$. This conformal factor satisfies the Lichnerowicz equation
\[ - \Delta_g \hat \phi - \frac{1}{4} \Lambda \hat \phi^5 = 0, \]
with the vanishing extrinsic curvature. Consequently, the metric $\hat g$ and the extrinsic curvature $\hat K_{ij} = 0$ already satisfy the constraint equations with the cosmological constant $\Lambda$. A coordinate transformation $r = \exp(\alpha)$ yields
\begin{equation}
\label{torus}
\hat g = \frac{1}{\Lambda} \left( d\alpha^2 + d \theta^2 + \sin^2 \theta d \varphi^2 \right).
\end{equation}
If we choose to identify $\alpha$ periodically with a period $T$, we get a round metric on $\mathbb S^1(T) \times \mathbb S^2$. The scalar curvature $\hat R$ of $\hat g$ is constant; $\hat R = 2 \Lambda$. 

The aim at this point is to construct an equivalent to the original Bowen--York expression (\ref{byrot}) on $\mathbb S^1(T) \times \mathbb S^2$ with the metric $\hat g$. This can also be done by exploiting the relations of the conformal method described above. Setting $\hat K^{ij} = \hat \phi^{-10} K^{ij}$, we get $\hat \nabla_i \hat K^{ij} = 0$, where $\hat \nabla_i$ denotes the covariant derivative with respect to the metric $\hat g_{ij}$. In spherical coordinates $(r,\theta,\phi)$ with the axis parallel to the angular momentum $J^i$ the only nonvanishing components of (\ref{byrot}) are $K_{r \varphi} = -3 J \sin^2 \theta/r^2$, where $J^2 = J_i J^i$. This leads to $\hat K_{\alpha \varphi} = \hat \phi^{-2} \frac{\partial r}{\partial \alpha} K_{r \varphi} = -3 J \sqrt{\Lambda} \sin^2 \theta$. A direct calculation yields, for the original Bowen--York extrinsic curvature (\ref{byrot}),
\[ K_{ij}K^{ij} = \frac{18}{r^6} \left[ J_i J^i - (J_i n^i)^2 \right], \]
or, in spherical coordinates defined above, $K_{ij} K^{i,j} = 18 \sin^2 \theta J^2/r^6$. Accordingly, $\hat K_{ij} \hat K^{ij} = 18 J^2 \Lambda^3 \sin^2 \theta = 8 b^2 \Lambda \sin^2 \theta$, where $b = 3 J \Lambda / 2$. The initial data satisfying constraint equations (\ref{einsteinid}) can be obtained as $\tilde g_{ij} = \phi^4 \hat g_{ij}$, $\tilde K_{ij} = \phi^{-2} \hat K_{ij}$, provided that the conformal factor $\phi$ satisfies the equation
\[ - \Delta_{\hat g} \phi + \frac{1}{8} \hat R \phi - \frac{1}{4} \Lambda \phi^5 - \frac{\hat K_{ij} \hat K^{ij}}{8 \phi^7} = 0, \]
where $\Delta_{\hat g}$ denotes the Laplacian with respect to metric $\hat g_{ij}$. In explicit terms we get
\begin{equation}
\label{lichnerowicz}
- \partial_{\alpha \alpha} \phi - \frac{1}{\sin \theta} \partial_\theta \left( \sin \theta \partial_\theta \phi \right) - \frac{1}{\sin^2 \theta} \partial_{\varphi \varphi} \phi + \frac{1}{4} \phi - \frac{1}{4} \phi^5 - \frac{b^2 \sin^2 \theta}{\phi^7} = 0.
\end{equation}
Here $\alpha \in \mathbb S^1(T)$, $(\theta, \varphi) \in \mathbb S^2$; by $\alpha \in \mathbb S^1(T)$ we mean that the solution of Eq.\ (\ref{lichnerowicz}) should be periodic in $\alpha$ with the period $T$. This article is devoted to the analysis of solutions of Eq.\ (\ref{lichnerowicz}), their dependence on the parameter $b$, and bifurcations.

The remaining sections are organized as follows. Section \ref{sec_sols} is devoted to the analysis of solutions of Eq.\ (\ref{lichnerowicz}). In subsection \ref{sec_bifurcation} we discuss the bifurcation pattern of solutions, and collect basic existence results. A detailed account of these and related facts can be found in \cite{BPS}. In subsection \ref{bzero} we deal with the exactly solvable case of Eq.\ (\ref{lichnerowicz}) with $b = 0$. In subsection \ref{general_method} we describe the general numerical spectral method used to obtain solutions for $b \neq 0$.

Section \ref{linear} is dedicated to the analysis of the linearized Eq.\ (\ref{lichnerowicz}). The main reason of presenting this analysis here is that it is directly related to the bifurcation structure of the solutions. Best known theorems formalizing this relation were proved by Crandall and Rabinowitz \cite{crandall_rabinowitz1, crandall_rabinowitz2}. Roughly speaking, they connect the bifurcation of solutions with an occurrence of a zero mode in the linearized equation. In the context of Eq.\ (\ref{lichnerowicz}), the analysis of the corresponding linearized equation clarifies the bifurcation structure described in subsection \ref{sec_bifurcation}, but it can also be used as an evidence suggesting that no additional branches of solutions bifurcate from the already known ones. Strictly speaking, the theorems of Crandall and Rabinowitz \cite{crandall_rabinowitz1, crandall_rabinowitz2} apply to bifurcations from simple zero eigenvalues. We will see in Sec.\ \ref{linear} that in our case bifurcations occur for degenerate zero eigenvalues. A generalization of theorems of Crandall and Rabinowitz that works in the case investigated in this paper was given by Vanderbauwhede \cite{vanderbauwhede}. Vanderbauwhede's theorem requires that a background solution (a solution from which a new branch bifurcates) shares a continuous symmetry --- the $O(2)$ symmetry in our case. We summarize the results on existence of the symmetry-breaking bifurcations for the solutions of Eq.\ \ref{lichnerowicz} in Corollary \ref{cor1}. Section \ref{linear} is also divided into subsections. In subsection \ref{separation} we discuss the separation of variables in the linearized equation. The case with $b = 0$ is then treated in subsection \ref{linb0}. The general spectral numerical method used to solve the linear eigenvalue problem is described in subsection \ref{lin_general_method}. The numerical results are discussed in subsection \ref{lin_results}. Section \ref{conclusions} contains a few concluding remarks.

\section{Solutions of the Lichnerowicz equation}
\label{sec_sols}

\subsection{Bifurcation diagrams}
\label{sec_bifurcation}

\begin{figure}[t]
\begin{center}
\includegraphics[width=0.8\textwidth]{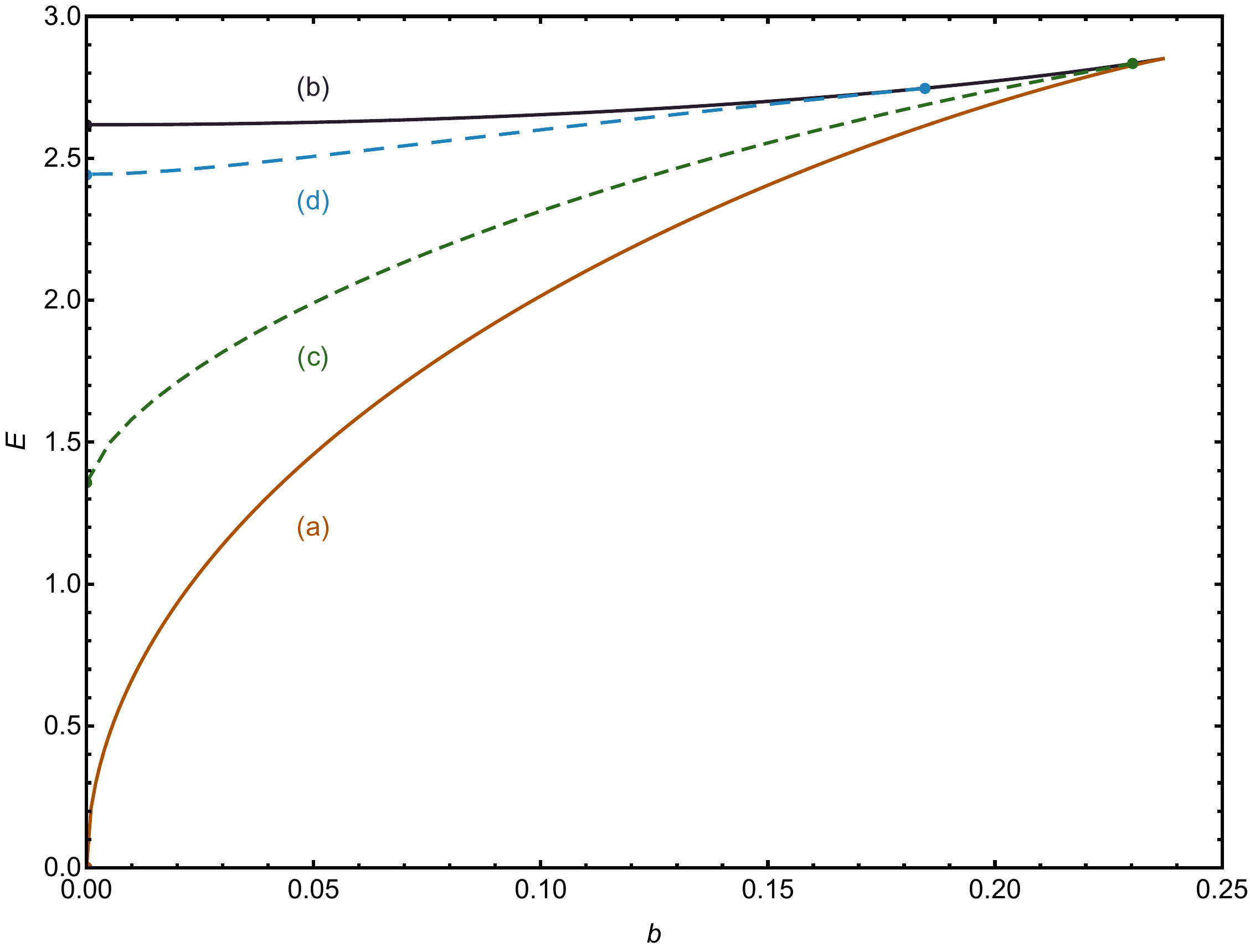}
\end{center}
\caption{\label{fig2}Energies of different solutions for $T = 5\pi$.}
\end{figure}

\begin{figure}[t]
\begin{center}
\includegraphics[width=0.8\textwidth]{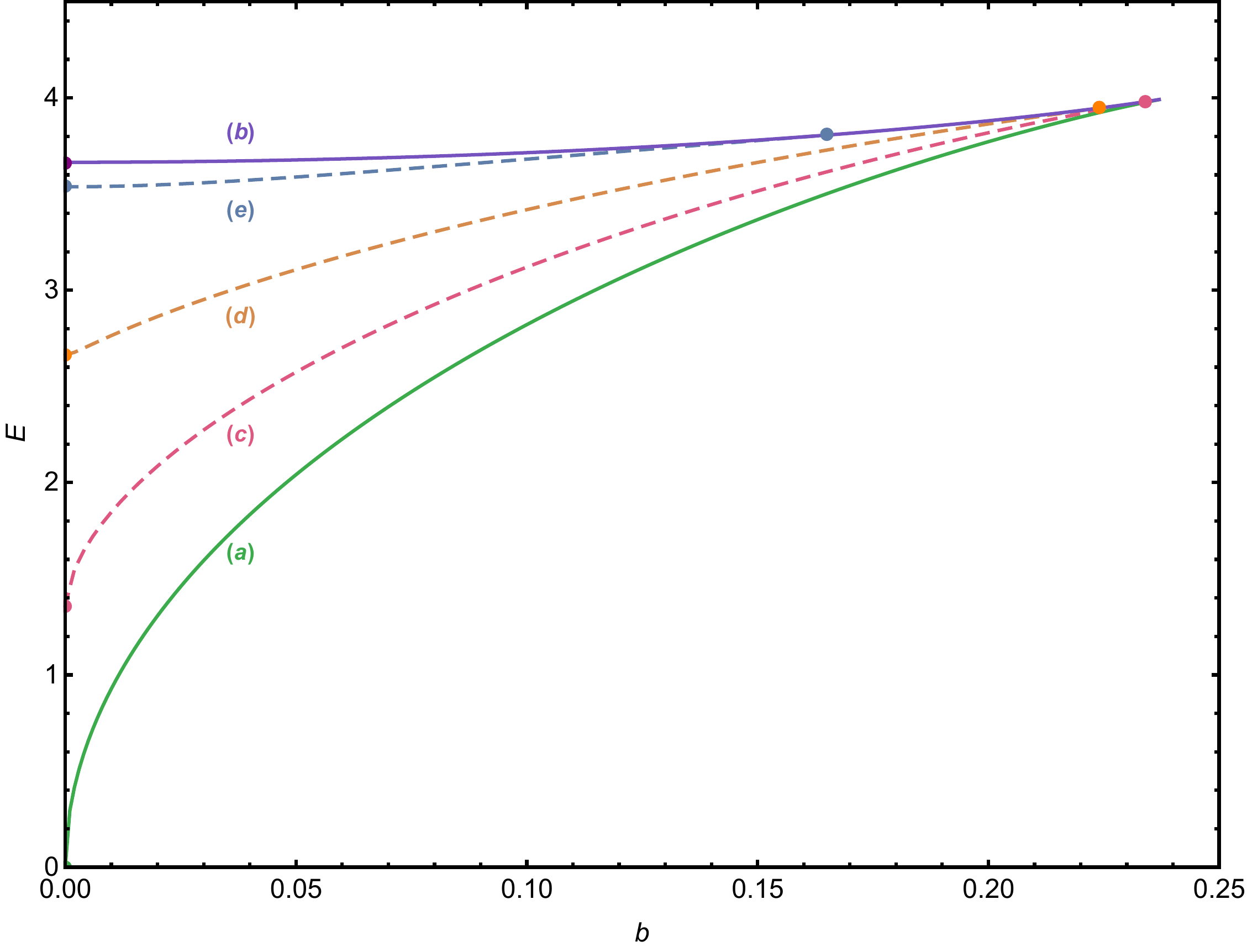}
\end{center}
\caption{\label{bifdiag}Same as in Fig.\ \ref{fig2}, but for $T = 7\pi$.}
\end{figure}

We begin our discussion with an example of a family of solutions of Eq.\ (\ref{lichnerowicz}) obtained numerically for $T = 5 \pi$. For conciseness we will restrict most of our numerical examples to this sample period. Our numerical solutions do not depend on $\varphi$, i.e., they all admit the $O(2)$ symmetry group acting on $\mathbb S^2$. Figure \ref{fig2} shows a bifurcation diagram created by computing the value of the `energy' functional
\[ E = \int_0^T d\alpha \int_0^\pi \sin \theta d \theta \left[ \frac{1}{2} (\partial_\alpha \phi)^2 + \frac{1}{2} (\partial_\theta \phi)^2 + \frac{1}{8} \phi^2 - \frac{1}{24} \phi^6 + \frac{1}{6} \frac{b^2 \sin^2 \theta}{\phi^6} \right] \]
associated with each of the solutions depending on $\alpha$ and $\theta$. Note that the `full' energy associated with solutions that would potentially depend also on $\varphi$, and that corresponds directly to Eq.\ (\ref{lichnerowicz}), reads
\begin{eqnarray*}
I & = & \int_0^T d\alpha \int_0^\pi \sin \theta d \theta \int_0^{2 \pi} d \varphi \left[ \frac{1}{2} (\partial_\alpha \phi)^2 + \frac{1}{2} (\partial_\theta \phi)^2 + \frac{1}{2} \frac{1}{\sin^2 \theta} (\partial_\varphi \phi)^2 \right. \\
& & \left. + \frac{1}{8} \phi^2 - \frac{1}{24} \phi^6 + \frac{1}{6} \frac{b^2 \sin^2 \theta}{\phi^6} \right].
\end{eqnarray*}
For the solutions that do not depend on $\varphi$ one has $I = 2 \pi E$. The abscissa in Fig.\ \ref{fig2} shows the bifurcation parameter $b$. 

There are four branches of solutions for $T = 5\pi$: two branches consisting of solutions that only depend on $\theta$ [admitting the $O(2) \times O(2)$ symmetry group acting on $\mathbb S \times \mathbb S^2$; they are denoted as branch (a) and (b)], and two branches of solutions that do depend on both $\alpha$ and $\theta$ [branch (c) and (d)]. The solutions exist only for sufficiently small values of the parameter $b < b_\mathrm{max}$. This stays in agreement with a result obtained in \cite{BPS} that is based on a general theorem by Premoselli \cite{premoselli}, and which we also quote at the end of this section. There are four solutions for $b = 0$, each belonging to one of the branches (a--d). Two of these solutions are constant: $\phi \equiv 1$ ($E = T/6$) and $\phi \equiv 0$ ($E = 0$). The remaining two are non trivial, but they only depend on $\alpha$. They are depicted in Fig.\ \ref{fig1}.  We derive these solutions in Sec.\ \ref{bzero}. 

The $O(2) \times O(2)$-symmetric branches (a) and (b) were obtained numerically already in \cite{BPS}. They originate at $b = 0$ from the two elementary solutions $\phi \equiv 0$ [branch (a)] and $\phi \equiv 1$ [branch (b)], and they join each other as $b \to b_\mathrm{max}$. The existence of symmetry-breaking of branches (c) and (d) was suggested in \cite{BPS}. They bifurcate from branch (b). As $b \to 0$, these branches join the two nontrivial solutions that depend on $\alpha$ only, and that were plotted in Fig.\ \ref{fig1}. We refer to the branch that bifurcates at $b = b_1 = 0.235$ as branch (c) and to the one bifurcating at $b = b_2 = 0.188$ as branch (d). Sample solutions belonging to these two branches are shown in Figs.\ \ref{fig3} and \ref{fig4}.

An example of a bifurcation diagram analogous to that of Fig.\ \ref{fig2}, but obtained for $T = 7\pi$ is shown in Fig.\ \ref{bifdiag}. There are five branches of solutions: two $O(2) \times O(2)$ symmetric branches (a) and (b), and three symmetry-breaking branches (c), (d) and (e).  In general for a given period $T$, we expect $k$ symmetry-breaking branches, where $k$ is the largest integer satisfying $k < T/(2\pi)$. We will return to this point in Corollary \ref{cor1}.

Except for the case with $b = 0$, all solutions discussed here are found numerically. There are, however, partial analytic existence, symmetry and stability results. They are described in detail in \cite{BPS}; here we only review those that are relevant for our discussion.

Denote the left-hand side of Eq.\ (\ref{lich1}) by
\[ F(\phi) = - \Delta_g \phi + \frac{1}{8}R \phi - \frac{1}{4} \Lambda \phi^5 - \frac{\Omega^2}{8 \phi^7}. \]
Stability of solutions of Eq.\ (\ref{lich1}) is understood in terms of the derivative
\[ F_\phi w = - \Delta_g w + \frac{1}{8} R w - \frac{5}{4} \Lambda \phi^4 w + \frac{7 \Omega^2}{8 \phi^8} w. \]
A solution $\phi$ of Eq.\ (\ref{lich1}) is called strictly stable, stable, marginally stable, unstable, or strictly unstable, if the lowest eigenvalue $\lambda$ of $F_\phi$ satisfies $\lambda > 0$, $\lambda \ge 0$, $\lambda = 0$, $\lambda \le 0$, or $\lambda < 0$, respectively. A motivation of this definition comes from the bifurcation theory, where the bifurcation points are identified by zero eigenvalues of $F_\phi$. In particular, it is not connected with the dynamical stability of solutions to the Einstein equations with the initial data implied by a given conformal factor $\phi$.

The following important facts are proved/listed in \cite{BPS}.

\begin{theorem}[Bizo\'{n}, Pletka, Simon]
Let $(\mathcal{M},g_{ij},K_{ij})$ be a seed manifold such that $g_{ij}$ and $\Omega^2 = K_{ij}K^{ij}$ admit a continuous symmetry $\xi$, i.e.,
\[ \mathcal{L}_\xi g_{ij} = 0, \quad \mathcal{L}_\xi \Omega = 0, \]
where $\mathcal{L}$ is the Lie derivative. Then all stable solutions of (\ref{lich1}) are also symmetric, that is
\[ \mathcal{L}_\xi \phi = 0. \]
\end{theorem}

The next result by Bizo\'{n}, Pletka and Simon is based on the work of Hebey, Pacard and Pollack \cite{hebey_pacard_pollack}. The proof can be found in \cite{pletka thesis}.

\begin{theorem}[Hebey, Pacard, Pollack; Bizo\'{n}, Pletka, Simon]
Consider a seed manifold $(\mathcal{M}, g_{ij}, K_{ij})$ as defined in the Introduction, but with a constant scalar curvature $R$.
\begin{enumerate}
\item Let
\[ \int_\mathcal{M} \Omega^2 dV \le \frac{Y^6}{256\Lambda^2 R^3 V^3}, \]
then Eq.\ (\ref{lich1}) has a smooth positive solution.
\item Assume that
\[ \int_\mathcal{M} \Omega^{5/6} dV > \frac{R^{5/4}V}{3^{5/4} \Lambda^{5/6}}, \]
then Eq.\ (\ref{lich1}) has no smooth positive solution.
\end{enumerate}
Here $V = \int_\mathcal{M}dV$ is the volume of $\mathcal{M}$, $dV$ is the volume element associated with the metric $g_{ij}$, and
\[ Y = \inf_{\gamma \in C^\infty(\mathcal{M}), \, \gamma \not\equiv 0} \frac{\int_\mathcal{M} (8 |\nabla \gamma|^2 + R \gamma^2) dV}{\left( \int_\mathcal{M} \gamma^6 dV \right)^{1/3}}. \]
is the Yamabe constant.
\end{theorem}

Another important result was obtained by Premoselli \cite{premoselli}.

\begin{theorem}[Premoselli]
\label{thm_prem}
Let us decompose $\Omega= \tilde b \Omega_0$ in (\ref{lich1}) in terms of a constant $\tilde b$ and a fixed function $\Omega_0$. There exists $0 < b_\mathrm{max} < \infty$ such that Eq.\ (\ref{lich1}) has
\begin{enumerate}
\item At least two positive solutions for $\tilde b < b_\mathrm{max}$, at least one of which is strictly stable. In addition, one of the strictly stable solutions, called $\phi(\tilde b)$, is `minimal' --- for any positive solution $\phi \not\equiv \phi(\tilde b)$ one has $\phi > \phi(\tilde b)$.
\item A unique, marginally stable, positive solution for $\tilde b = b_\mathrm{max}$.
\item No solution for $\tilde b > b_\mathrm{max}$.
\end{enumerate}
\end{theorem}

The decomposition $\Omega= \tilde b \Omega_0$ is, of course, non-unique. A natural choice in case of Eq.\ (\ref{lichnerowicz}) is to assume $\tilde b = b = 3 J \Lambda / 2$.

The branch of stable solutions whose existence is asserted by the above theorem can be identified with the one that starts from $\phi \equiv 0$ at $b = 0$, which we consequently denote as branch (a) in Figs.\ \ref{fig2}, \ref{bifdiag}.

\subsection{$O(3)$-symmetric solutions ($b = 0$)}
\label{bzero}

\begin{figure}[t]
\begin{center}
\includegraphics[width=0.8\textwidth]{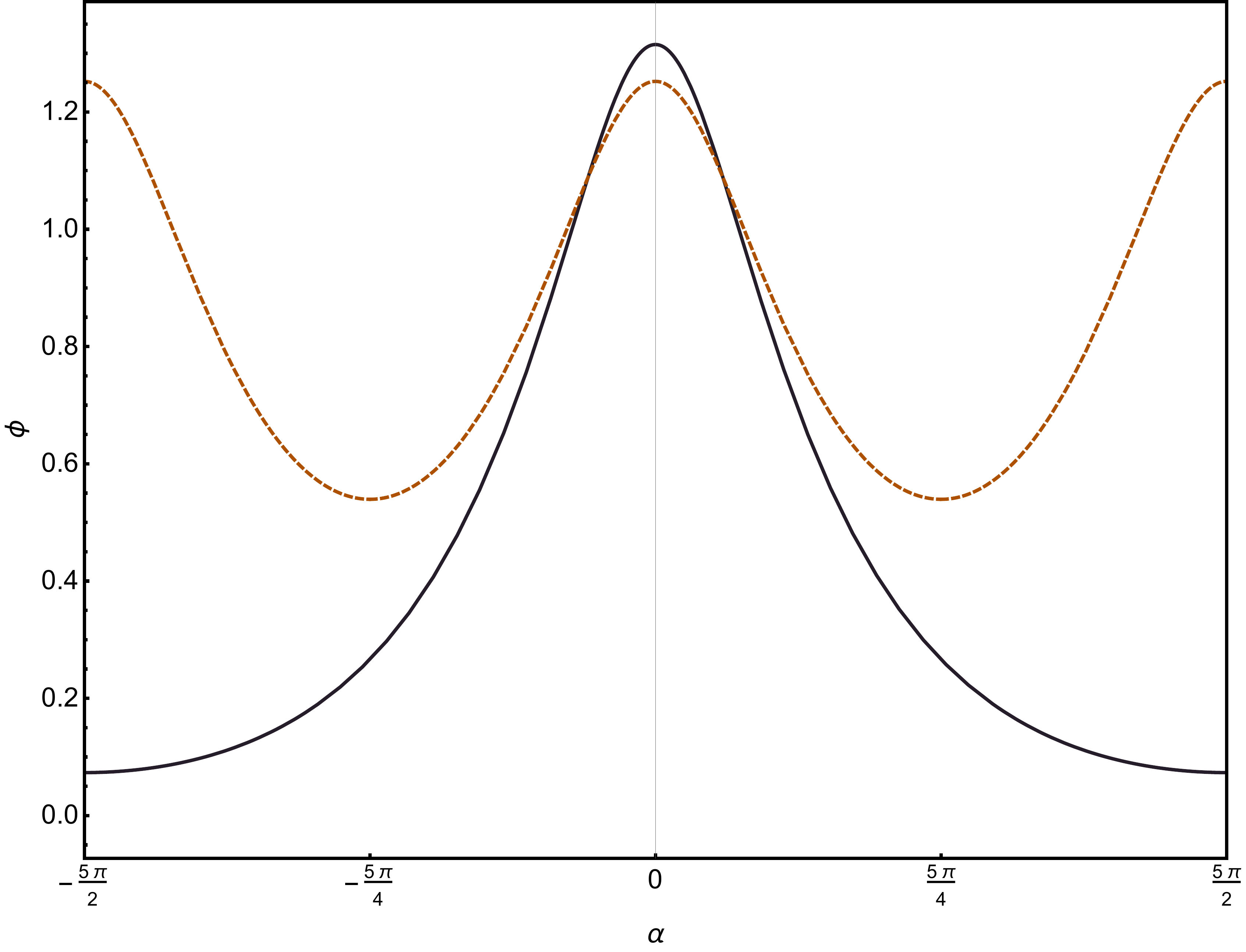}
\end{center}
\caption{\label{fig1}Periodic solutions with period $T = 5\pi$ and $b = 0$.}
\end{figure}

For $b = 0$ Eq.\ (\ref{lichnerowicz}) admits solutions that do not depend on $\theta$ and $\varphi$. In this case it can be written as an ordinary differential equation
\begin{equation}
\label{lane_emden_a}
\frac{d^2 \phi}{d\alpha^2} = \frac{1}{4} \phi (1 - \phi^4),
\end{equation}
where $\phi$ should be a periodic function of $\alpha$.

Two obvious solutions are $\phi \equiv 0$ and $\phi \equiv 1$. The existence of the solution $\phi \equiv 1$ is a consequence of the fact that metric (\ref{torus}) satisfies the constraint equations (\ref{einsteinid}) itself. It represents a time-symmetric slice through the Nariai spacetime \cite{nariai}
\[ g = \frac{1}{\Lambda}(-dt^2 + \cosh^2 t d\alpha + d\theta^2 + \sin^2 \theta d\varphi^2). \]

Remarkably, Eq.\ (\ref{lane_emden_a}) corresponds to the famous Lane--Emden equation with index $n = 5$,
\begin{equation}
\label{lane_emden_b}
\frac{1}{r^2} \frac{d}{dr} \left( r^2 \frac{d \Phi}{dr} \right) + \Phi^5  = 0,
\end{equation}
where $r$ is a spherical radius. The relation between the solutions of Eqs.\ (\ref{lane_emden_a}) and (\ref{lane_emden_b}) is given by $\Phi = \phi/\sqrt{2r}$, $\alpha = - \ln r$.
All real solutions of Eqs.\ (\ref{lane_emden_a}) and (\ref{lane_emden_b}) are derived in \cite{mach}. Here we are interested only in those that are periodic and everywhere positive. Multiplying Eq.\ (\ref{lane_emden_a}) by $d\phi/d\alpha$ and integrating with respect to $\alpha$ one obtains
\[ \left( \frac{d\phi}{d\alpha} \right)^2 = \frac{1}{12} \left( -\phi^6 + 3 \phi^2 + C \right), \]
where $C$ is an integration constant. Positive and periodic solutions exist for $C \in (-2,0)$. In this case the polynomial $w(\phi) =  -\phi^6 + 3 \phi^2 + C$ has 4 real roots. It can be factorized as
\[ w(\phi) = (\phi^2 - a)(b - \phi^2)(\phi^2 + c),  \]
where
\begin{eqnarray*}
a & = & 2 \sin \left[ \frac{1}{3} \mathrm{arc\, sin} \left( \frac{|C|}{2} \right) \right], \quad b = 2 \cos \left[ \frac{1}{3} \mathrm{arc\, cos} \left( - \frac{|C|}{2} \right) \right], \\
c & = & 2 \cos \left[ \frac{1}{3} \mathrm{arc\, cos} \left( \frac{|C|}{2} \right) \right].
\end{eqnarray*}
It is elementary to show that
\[ 0 < a < 1 < b < \sqrt{3} < c < 2. \]
Solutions of Eq.\ (\ref{lane_emden_a}) can be then written as
\begin{equation}
\label{sol1}
\phi = \sqrt{\frac{a b y^2}{b y^2 - (b - a)}}, \quad y = \mathrm{dc} \left[ \frac{1}{2} \sqrt{\frac{(a+c)b}{3}} (\alpha - \alpha_0) , \sqrt{\frac{(b - a)c}{(a + c)b}} \right],
\end{equation}
where $\mathrm{dc}$ is a subsidiary Jacobian elliptic function (in the standard Glaisher notation) \cite{mach}. Note that $|\mathrm{dc}(x,k)| \ge 1$ for $k \in [0,1]$. It follows that the solution (\ref{sol1}) is strictly positive.

The function $\mathrm{dc}(x,k)$ is periodic with the period $4K(k)$, where
\[ K(k) = \int_0^\frac{\pi}{2} \frac{d \theta}{\sqrt{1 - k^2 \sin^2 \theta}} \]
is the complete elliptic integral of the first kind. The period of $y$ with respect to $\alpha$ is
\[ 2 T_\phi \equiv 8 \sqrt{\frac{3}{(a+c)b}}  K\left( \sqrt{\frac{(b - a)c}{(a + c)b}} \right), \]
and because $\phi$ depends on $y^2$, it is periodic with the period $T_\phi$. The period $T_\phi$ given by the above formula is a strictly increasing function of $C \in [-2,0)$. It is equal to $2 \pi$ for $C = -2$ and tends to $+ \infty$, as $C \to 0$. Finding the value of $C$ corresponding to a given period $T_\phi$ is a simple numerical task. In Figure \ref{fig1} we plot two sample solutions $\phi(\alpha)$ corresponding to the period $T = 5\pi$. They are given by Eq.\ (\ref{sol1}) for $T_\phi = 5\pi$ (in this case $C = -0.0161381$) and $T_\phi = 5 \pi/2$ ($C = -0.848422$).

In general, for a given period $T$, there are $k$ periodic solutions, where $k$ is the largest integer satisfying $k < T/(2 \pi)$. The periods of these solutions read $T_\phi = T/j$, where $j \le k$. It is easy to show that they correspond to time-symmetric initial data obtained by gluing together $j$ copies of the $t = \mathrm{const}$ slices through the Kottler (Schwarzschild--de Sitter) metric
\[ g = - \left(1 - \frac{2M}{r} - \frac{\Lambda}{3}r^2 \right) dt^2 + \frac{dr^2}{1 - \frac{2M}{r} - \frac{\Lambda}{3}r^2} + r^2 (d \theta^2 + \sin^2 \theta d\phi^2). \]
Such initial data are periodic in the radial coordinate and contain $j$ pairs of horizons --- each pair consists of a `cosmological' and a `black hole' horizon. The constant $C$ is related with the Kottler mass by $C = - 6 M \sqrt{\Lambda}$. Showing this amounts to a simple modification of a calculation done in \cite{mach_niall}. It is expected that each of these periodic solutions belongs to one of symmetry-breaking branches discussed in Sec.\ \ref{sec_bifurcation}.

For $C = -2$ one gets a solution of the form $\phi \equiv 1$. The other limiting case, with $C = 0$, yields the well known Schuster solution of the Lane--Emden equation
\[ \Phi(r) = \frac{1}{\sqrt{1 + \frac{1}{3}r^2}}. \]
In terms of $\phi$ and $\alpha$ it is given by
\begin{equation}
\label{sol2}
\phi = \left( \frac{12}{4} \right)^\frac{1}{4} \sqrt{\mathrm{sech} (\alpha - \alpha_0)},
\end{equation}
which may be interpreted as a case with infinite period.

\subsection{$O(2)$-symmetric solutions}
\label{general_method}

\begin{figure}[t]
\begin{center}
\includegraphics[width=0.8\textwidth]{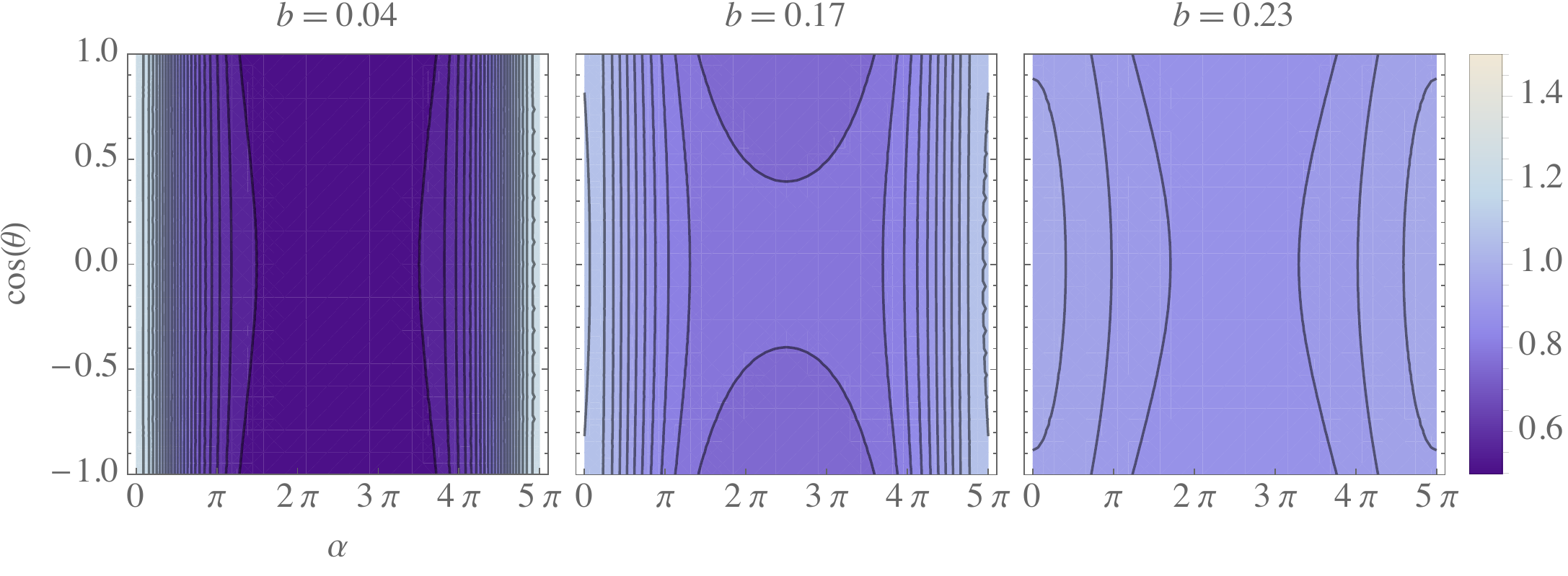}
\end{center}
\caption{\label{fig4}Contour plots of sample solutions depending on $\alpha$ and $\theta$. Here $T = 5\pi$. The above solutions belong to branch (c) plotted in Fig.\ \ref{fig2}.}
\end{figure}

\begin{figure}[t]
\begin{center}
\includegraphics[width=0.8\textwidth]{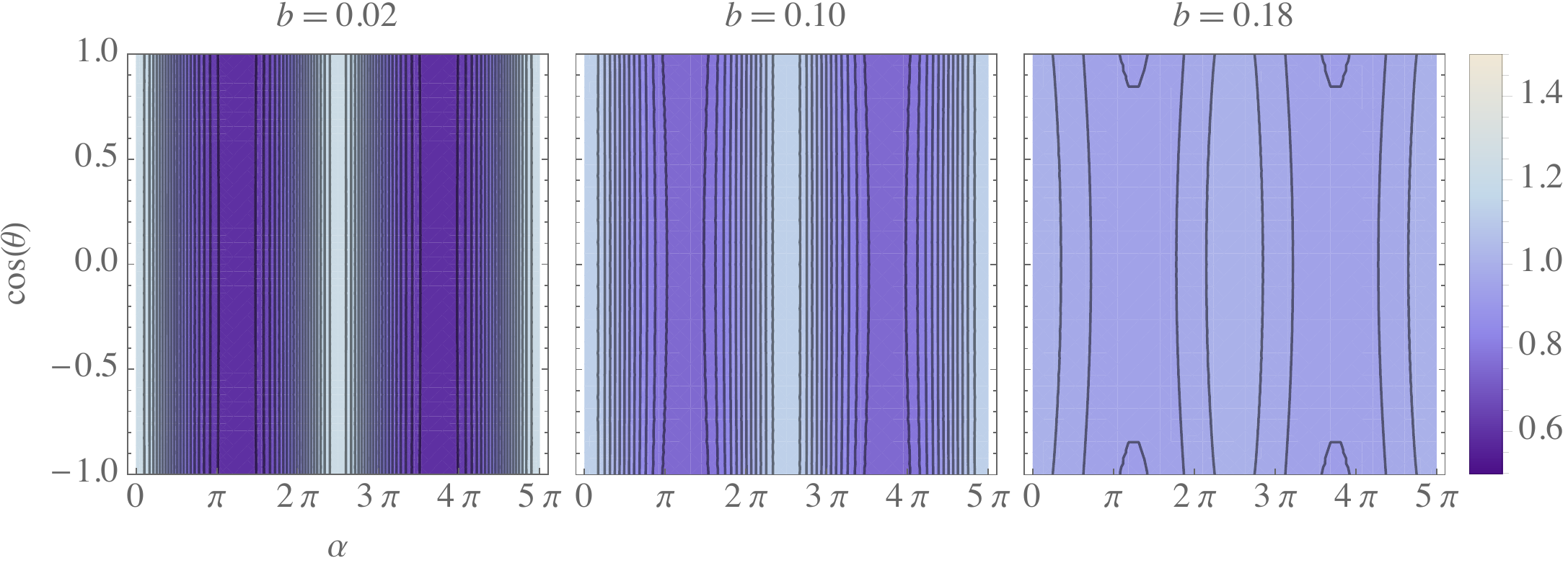}
\end{center}
\caption{\label{fig3}Contour plots of sample solutions depending on $\alpha$ and $\theta$. Here $T = 5\pi$. The above solutions belong to branch (d) in Fig.\ \ref{fig2}.}
\end{figure}

In the more general case of solutions depending both on $\alpha$ and $\theta$, but not on $\varphi$,  we resort to numerical methods. In this case Eq.\ (\ref{lichnerowicz}) can be written as
\begin{equation}
\label{lichnerowicz2}
- \partial_{\alpha \alpha} \phi - \frac{1}{\sin \theta} \partial_\theta \left( \sin \theta \partial_\theta \phi \right) + \frac{1}{4} \phi - \frac{1}{4} \phi^5 - \frac{b^2 \sin^2 \theta}{\phi^7} = 0.
\end{equation}
We solve this equation numerically using a spectral scheme, which we now describe briefly.

The solution is approximated as
\begin{equation}
\label{expansion}
\tilde \phi (\alpha, \theta) = \sum_{k = 0}^M \sum_{l=0}^N a_{kl} y_{kl}(\alpha, \theta),
\end{equation}
where the expansion functions $y_{kl}$, $k = 0, \dots, M$, $l = 0, \dots, N$, are eigenfunctions of the operator $L$ defined as
\[ L \phi \equiv - \partial_{\alpha \alpha} \phi - \frac{1}{\sin \theta} \partial_\theta \left( \sin \theta \partial_\theta \phi \right) + \frac{1}{4} \phi \]
[$L \phi$ is the linear part of the left-hand side of Eq.\ (\ref{lichnerowicz2})]. We choose specifically
\begin{eqnarray}
\label{ya}
y_{0l}(\alpha, \theta) & = & \frac{1}{2} P_l(\cos \theta), \quad l = 0, \dots, N, \\
y_{kl}(\alpha, \theta) & = & \cos \left( k \frac{2 \pi}{T} \alpha \right) P_l (\cos \theta), \quad k = 1, \dots, M, \quad l = 0, \dots, N,
\label{yb}
\end{eqnarray}
where $P_l$ denotes the $l$-th Legendre polynomial. The corresponding eigenvalues of $L$ read
\[ L y_{kl} = \left[  \left( \frac{2 \pi}{T} \right)^2 k^2 + l(l+1) + \frac{1}{4} \right] y_{kl}. \]
Since the eigenfunctions $u_{kl}$ satisfy the periodic boundary conditions, so does the solution $\tilde \phi$. This approach is sometimes referred to as the Galerkin method.

The nonlinear part in Eq.\ (\ref{lichnerowicz2}) is more troublesome. The task is to expand the expression
\[ -\frac{1}{4} \tilde \phi^5 - \frac{b^2 \sin^2 \theta}{\tilde \phi^7} \]
in $y_{kl}$. The coefficients of this expansion are given by the integrals
\begin{eqnarray}
\nonumber
c_{kl} & = & \frac{2l+1}{T} \int_0^\pi d\theta \sin \theta \int_0^T d\alpha \left( - \frac{1}{4} \tilde \phi^5 - \frac{b^2 \sin^2 \theta}{\tilde \phi^7} \right) \\
& & \times \cos\left( k \frac{2\pi}{T} \alpha \right) P_l(\cos \theta).
\label{bkl}
\end{eqnarray}
The values of $c_{kl}$ are computed using Gauss--Legendre--Fourier quadratures. We describe this procedure in the Appendix.

Equation (\ref{lichnerowicz2}) yields the following set of equations for the coefficients $a_{kl}$
\begin{eqnarray*}
& & r_{kl} \equiv \left[ \left( \frac{2 \pi}{T} \right)^2 k^2 + l(l+1) + \frac{1}{4} \right] a_{kl} + c_{kl} = 0, \\
& & \quad k = 0, \dots, M, \quad l = 0, \dots, N.
\end{eqnarray*}
Note that the coefficients $c_{kl}$ depend on $a_{kl}$ via Eqs.\ (\ref{expansion}) and (\ref{bkl}). We solve the above set of equations iteratively, using the Newton--Raphson scheme. In each iteration corrections $\Delta_{kl}$ to the coefficients $a_{kl}$ are computed as a solution of the set of linear equations
\[ \sum_{m=0}^M \sum_{n=0}^N \frac{\partial r_{kl}}{\partial a_{mn}} \Delta_{mn} = - r_{kl}, \quad k = 0, \dots, M, \quad l = 0, \dots, N, \]
where the Jacobians $\partial r_{kl}/\partial a_{mn}$ are computed numerically. A new set of coefficients $a^\prime_{kl}$ is then given by $a^\prime_{kl} = a_{kl} + \Delta_{kl}$. It turns out that the above method converges relatively fast. Around 10 to 15 iterations suffice to converge up to machine precision. The numerical error is thus controlled by $N$, $M$, and the accuracy with which the Jacobians $\partial r_{kl}/\partial a_{mn}$ are computed.

Sample solutions obtained with the above method are shown in Figs.\ \ref{fig4} and \ref{fig3}. They belong to branches (c) and (d) illustrated in Fig.\ \ref{fig2} and discussed in previous sections. It is worth noticing that both sequences of solutions join smoothly with the solutions depending only on $\alpha$ for $b = 0$ and those depending only on $\theta$ for $b = b_1$ and $b_2$, respectively.

\section{Linearized eigenvalue problem}
\label{linear}

In this section we prove existence of bifurcation points giving rise to the symmetry-breaking branches of solutions depending on $\alpha$ and $\theta$. We are also concerned with the question of existence or nonexistence of additional solutions of Eq.\ (\ref{lichnerowicz}). This general problem can be naturally divided into following specific cases or tasks:
\begin{enumerate}
\item Symmetry of solutions with respect to $\varphi$. Is it possible to prove that there are no solutions depending on $\varphi$?
\item A weaker version of the above question: Is it possible to prove that no solution that depends on $\varphi$ bifurcate from the known solutions, e.g.\ those depicted in Fig.\ \ref{fig2}?
\item Are there any other solutions depending on $\alpha$ and $\theta$ only that do not belong to the already known branches of solutions?
\item Analogously: can one prove that no branch of solutions depending only on $\alpha$ and $\theta$ bifurcates from the known branches?
\end{enumerate}
We believe that it might be possible to answer the first of the above questions by means of techniques that stem from the method of moving planes \cite{ni_nirenberg}, including in particular a version that was called the method of moving spheres \cite{jin}. In this manner Chru\'{s}ciel and Gicqaud were able to show that solutions of an equation similar to Eq.\ (\ref{lichnerowicz}) are symmetric with respect to $\varphi$ \cite{chrusciel}.

In this paper we collect numerical evidence concerning questions (ii) and (iv) --- that no additional branches of solutions bifurcate from the already known ones. The reasoning is based on the linearization of Eq.\ (\ref{lichnerowicz}), which is also a main tool allowing one to understand the bifurcation structure of its solutions. We exploit the fact that new branches of solutions bifurcate from solutions for which zero modes occur in the linearized equation.

It is well known that an existence of a zero eigenvalue of the linearized equation is not a sufficient condition for the occurrence of bifurcation. The best known rigorous result assuring the occurrence of bifurcation from simple zero eigenvalues is due to Crandall and Rabinowitz \cite{crandall_rabinowitz1, crandall_rabinowitz2}. We will see, however, that in our case bifurcations do occur from degenerate zero modes. A suitable generalization of the original theorem of Crandall and Rabinowitz was given by Vanderbauwhede \cite[Thm.\ 6.2.6]{vanderbauwhede}; it applies to potentially degenerate zero eigenvalues for which the eigenspace is invariant under the action of the orthogonal group $O(n)$. Below we quote this theorem after \cite{smoller_wasserman}.

\begin{theorem}[Vanderbauwhede]\label{thm4}
Let $X$, $B$ and $Z$ be real Banach spaces, and let $Y \subset X \times B$ be a neighborhood of the origin $(0,0) \in Y$. Consider a mapping $F \colon Y \to Z$ of class $C^2$ such that $F(0,0) = 0$. Assume further that the following hyphoteses are satisfied.
\begin{enumerate}
\item There exist representations of the orthogonal group $O(n)$ on $X$ and $Z$ denoted by $\Gamma \colon O(n) \to GL(X)$ and $\tilde \Gamma \colon O(n) \to GL(Z)$, where $GL(X)$ and $GL(Z)$ denote the general linear groups on $X$ and $Z$, respectively. Moreover, for each $s \in O(n)$ and all $(x,b) \in Y$, the following conditions hold:
\begin{enumerate}
\item $(\Gamma(s) x, b) \in Y$, and
\item $F(\Gamma(s)x,b) = \tilde \Gamma(s) F(x,b)$.
\end{enumerate}
\item The partial derivative $F_x(0,0)$ is a Fredholm operator of index zero, and the representation $\Gamma_0$ of $O(n)$ induced by $\Gamma$ on the kernel $\mathrm{ker} \, F_x(0,0)$ is irreducible.
\item There is a non-zero vector $u_0$ in $\mathrm{ker} \, F_x(0,0)$ such that $F_{xb}(0,0)(u_0,0) \notin \mathrm{Range}\,(F_x(0,0))$.
\end{enumerate}
Then $(0,0)$ is a bifurcation point of the equation $F(x,b) = 0$.
\end{theorem}

The above theorem applies to our case as follows. We define
\[ F(\phi, b^2) = - \Delta_{\hat g} \phi + \frac{1}{4} \phi - \frac{1}{4} \phi^5 - \frac{b^2 \sin^2 \theta}{\phi^7}, \]
so that Eq.\ (\ref{lichnerowicz}) can be written simply as $F(\phi,b^2) = 0$. Instead of the origin, we consider a point $(\phi_0,b_0^2)$ (which, of course, can be trivially shifted to the origin). The derivative $F_\phi$ at $(\phi_0,b_0^2)$ reads
\[ F_\phi (\phi_0,b_0^2) w = - \Delta_{\hat g} w + \frac{1}{4} w - \frac{5}{4} \phi_0^4  w + \frac{7 b^2 \sin^2 \theta}{\phi_0^8} w. \]
Clearly, $F_\phi$ is a self-adjoint operator with respect to the standard $L^2$ inner product on $\mathbb S^1(T) \times \mathbb S^2$, and thus it has a zero Fredholm index (i.e., the dimension of its kernel is the same, as the co-dimension of its range). Suppose now that $F_\phi (\phi_0,b_0^2)$ admits a zero eigenvalue. To see that the condition (iii) is satisfied, note that the mixed derivative $F_{\phi b^2}(\phi_0,b_0^2)$ reads
\[ F_{\phi b^2}(\phi_0,b_0^2) w = \frac{7 \sin^2 \theta}{\phi_0^8} w. \]
Let $w_0$ be a vector from $\mathrm{ker} \, F_\phi (\phi_0,b_0^2)$. We need to show that
\[ \frac{7 \sin^2 \theta}{\phi_0^8} w_0 \neq - \Delta_{\hat g} w + \frac{1}{4} w - \frac{5}{4} \phi_0^4  w + \frac{7 b^2 \sin^2 \theta}{\phi_0^8} w, \]
where $w$ is any vector in the domain of $F_\phi (\phi_0,b_0^2)$. To see that this condition indeed holds, assume the contrary, and multiply both sides of the obtained equation by $w_0$. We get
\begin{eqnarray*}
\frac{7 \sin^2 \theta}{\phi_0^8} w_0^2 & = & - w_0 \Delta w + \frac{1}{4} w_0 w - \frac{5}{4} \phi_0^4 w_0  w + \frac{7 b^2 \sin^2 \theta}{\phi_0^8} w_0 w \\
& = & - w_0 \Delta w + w \Delta w_0,
\end{eqnarray*}
where we have used the fact that $w_0 \in \mathrm{ker} \, F_\phi (\phi_0,b_0^2)$. Integrating over $\mathbb S(T) \times \mathbb S^2$ we get
\[ \int_0^T d\alpha \int_0^\pi d\theta \sin \theta \int_0^{2\pi} d \varphi \frac{7 \sin^2 \theta}{\phi_0^8} w_0^2 = 0, \]
which cannot hold, as the left-hand side of the above equation is strictly positive. In our case, bifurcations occur for the branch of solutions depending only on $\theta$ [denoted as branch (b) in Fig.\ \ref{fig2}], which are clearly symmetric with respect to translations in $\alpha$, i.e., they are $O(2)$ symmetric. This symmetry is broken by the bifurcating solutions [branches (c) and (d)]. We will see in the subsequent section that in our case $\mathrm{ker} \, F_\phi (\phi_0,b_0^2)$ is in fact two dimensional and that the representation of $O(2)$ induced on $\mathrm{ker} \, F_\phi (\phi_0,b_0^2)$ is irreducible.

In fact, Theorem \ref{thm4} and the above discussion imply the following result.

\begin{cor}\label{cor1}
Consider Eq.\ (\ref{lichnerowicz}) on $\mathbb S^1(T) \times \mathbb S^2$. Let the branch of solutions of Eq.\ (\ref{lichnerowicz}) which originates as $\phi \equiv 1$ for $b = 0$, and ends at $b = b_\mathrm{max}$, be denoted by (b), as in Figs.\ \ref{fig2} and \ref{bifdiag}. Let $k$ be the largest integer satisfying $k < T/(2\pi)$. Then there are $k$ bifurcation points on branch (b).
\end{cor}

The missing step in the proof of the above corollary, which we fill in below, is to show that the equation
\begin{equation}
\label{linf}
F_\phi (\phi,b^2) w = \lambda w
\end{equation}
has a zero eigenvalue at $k$ isolated points on branch (b).

In the following sections we will discuss particular cases in which the eigenvalue problem (\ref{linf}) can be solved exactly, or where the separation of variables can be carried out to some degree. We will then describe the numerical method which was used to compute the eigenvalues of Eq.\ (\ref{linf}) in more general cases.

\subsection{Linearization around solutions depending on $\alpha$ and $\theta$: separation of variables}
\label{separation}

Consider a solution $\phi$ of Eq.\ (\ref{lichnerowicz}) that does not depend on $\varphi$. The linearization of Eq.\ (\ref{lichnerowicz}) around $\phi$ leads to the following eigenvalue problem:
\begin{equation}
\label{schr}
 - \partial_{\alpha \alpha} w - \frac{1}{\sin \theta} \partial_\theta (\sin \theta \partial_\theta w) - \frac{1}{\sin^2 \theta} \partial_{\varphi \varphi}w + V(\alpha,\theta) w = \lambda w,
\end{equation}
where
\begin{equation}
\label{schr_potential}
V(\alpha,\theta) = \frac{1}{4} - \frac{5}{4} \phi^4(\alpha,\theta) + \frac{7 b^2 \sin^2 \theta}{\phi^8(\alpha,\theta)}.
\end{equation}
We require that the eigenfunctions $w \colon \mathbb S^1(T) \times \mathbb S^2 \to \mathbb R$ should be smooth on $\mathbb S^1(T) \times \mathbb S^2$.

The first immediate observation is that it is possible to factor out the dependence on $\varphi$. Assuming $w(\alpha, \theta, \varphi) = v(\alpha, \theta) z(\varphi)$ we obtain
\begin{equation}
\label{schr2}
 - \partial_{\alpha \alpha} v - \frac{1}{\sin \theta} \partial_\theta (\sin \theta \partial_\theta v) + \frac{m^2}{\sin^2 \theta}v + V(\alpha,\theta) v = \lambda v,
\end{equation}
where $V$ is given by Eq.\ (\ref{schr_potential}) and $m = 0, \pm 1, \pm 2, \dots$. Of course, one can also introduce a new potential $V_m(\alpha, \theta) = V(\alpha,\theta) + m^2/\sin^2 \theta$, and write
\begin{equation}
\label{schr3}
 - \partial_{\alpha \alpha} v - \frac{1}{\sin \theta} \partial_\theta (\sin \theta \partial_\theta v) + V_m(\alpha,\theta) v = \lambda v.
\end{equation}

\begin{remark}
Let $V_{min} = \min_{0 \le \alpha \le T, \; 0 \le \theta \le \pi} V(\alpha, \theta)$. Then $\lambda \ge V_{min}$.
\end{remark}
This is a well known fact. To obtain this estimate, it is enough to multiply both sides of Eq.\ (\ref{schr}) by $w$ and integrate with respect to $\int_0^T d \alpha \int_0^\pi d \theta \sin \theta \int_0^{2 \pi} d \varphi$. Integrating by parts one obtains
\begin{eqnarray*}
\lefteqn{\lambda \int_0^T d \alpha \int_0^\pi d \theta \sin \theta \int_0^{2 \pi} d \varphi w^2 = \int_0^T d \alpha \int_0^\pi d \theta \sin \theta \int_0^{2 \pi} d \varphi \left[ (\partial_\alpha w)^2  \right.} \\
& & \left. + (\partial_\theta w)^2 + \frac{1}{\sin^2 \theta} (\partial_\varphi w)^2 \right] + \int_0^T d \alpha \int_0^\pi d \theta \sin \theta \int_0^{2 \pi} d \varphi  w V w,
\end{eqnarray*}
from which the inequality follows immediately.

\begin{remark}
Applying an analogous reasoning to Eq.\ (\ref{schr2}) one gets an improved estimate $\lambda \ge V_{min} + m^2$ (note that $\min_{0 \le \theta \le \pi} m^2/\sin^2 \theta = m^2$).
\end{remark}

\begin{remark}
If $\phi$ depends on $\alpha$, the function $w = \partial_\alpha \phi$ is an eigenfunction of Eq.\ (\ref{schr}) corresponding to the eigenvalue $\lambda = 0$.
\end{remark}

To see this, it is enough to differentiate Eq.\ (\ref{lichnerowicz}) with respect to $\alpha$.

The following remark applies to Eq.\ (\ref{schr2}) or Eq.\ (\ref{schr3}).

\begin{remark}
\label{remark_separation}
If $\phi(\alpha, \theta) = \phi(\theta)$, then it makes sense to separate the variables as $v(\alpha,\theta) = v_1(\alpha) v_2(\theta)$.
\end{remark}

Inserting this ansatz into Eq.\ (\ref{schr2}) one obtains
\begin{equation}
\label{separation_theta}
\lambda = \lambda_\theta + \left( \frac{2\pi j}{T} \right)^2,
\end{equation}
where $j = 0, 1, \dots$, and $\lambda_\theta$ is an eigenvalue of the equation
\begin{equation}
\label{separation_eq}
- \frac{1}{\sin \theta} \frac{d}{d\theta} \left( \sin \theta \frac{d v_2}{d \theta} \right) + \frac{m^2}{\sin^2 \theta} v_2 + V(\theta) v_2 = \lambda_\theta v_2.
\end{equation}
The space $W$ of solutions $v_1$ is spanned by
\[ v_1 = A_j \sin(2 \pi j \alpha/T) + B_j \cos(2 \pi j \alpha/T), \quad j = 0, 1, \dots, \]
satisfying standard harmonic oscillator equation on $\mathbb S^1(T)$:
\[ \partial_{\alpha \alpha} v_1 + \left( \frac{2\pi j}{T} \right)^2 v_1 = 0. \]
Clearly the representation of the $O(2)$ group induced on $W$ is irreducible. Eigenvalues $2 \pi j/T$ are of course two-fold degenerate. Consequently, eigenvalues $\lambda$ are also degenerate.  

\subsection{Solutions for $b = 0$}
\label{linb0}

\subsubsection{Linearization around $\phi \equiv 0$ and $\phi \equiv 1$}

Equation (\ref{separation_eq}) can be solved in two limiting cases corresponding to $b = 0$ and $\phi \equiv 0$ or $\phi \equiv 1$, respectively. In both cases the potential $V$ is constant. We have $V \equiv -1$ for $\phi \equiv 1$ and $V \equiv 1/4$ for $\phi \equiv 0$. Defining $\mu = \cos \theta$ we can write Eq.\ (\ref{separation_eq}) as
\[ \frac{d}{d\mu}\left[ (1 - \mu^2) \frac{d}{d \mu} v_2 \right] + \left( \lambda_\theta - V - \frac{m^2}{1 - \mu^2} \right) v_2 = 0, \]
i.e., in the form of the general Legendre equation. It has regular solutions (associated Legendre polynomials $P^m_n(\mu)$), if and only if $\lambda_\theta - V = n(n+1)$, $n = 0, 1, \dots$, and $0 \le m \le n$ (or with equivalent negative values).

In summary, we get
\begin{equation}
\label{eigen_phi_1}
\lambda = - 1 + \left( \frac{2\pi j}{T} \right)^2 + n(n+1), \quad j,n = 0, 1, \dots,
\end{equation}
for $\phi \equiv 1$ and
\[ \lambda = \frac{1}{4} + \left( \frac{2\pi j}{T} \right)^2 + n(n+1), \quad j,n = 0, 1, \dots, \]
for $\phi \equiv 0$.

Clearly, for a given period $T$, there are $k + 1$ negative eigenvalues corresponding to $\phi \equiv 1$, where $k$ is the largest integer satisfying $k < T/(2\pi)$. The lowest eigenvalue is simply $\lambda_0 = -1$. The branch of solutions that originates at $\phi \equiv 1$, denoted as branch (b), ends at $b = b_\mathrm{max}$. On the other hand, we know from Theorem \ref{thm_prem} that the unique solution for $b = b_\mathrm{max}$ is marginally stable, i.e., the lowest eigenvalue corresponding to that solution is zero. It follows from continuity that there exist $k$ points in the interval $b \in (0, b_\mathrm{max})$, for which Eq.\ (\ref{schr2}) admits zero eigenvalues. Another way of seeing this follows directly from Eq.\ (\ref{separation_theta}), where the lowest eigenvalue $\lambda_\theta$ reads $\lambda_\theta = -1$ for $b = 0$ and $\lambda_\theta = 0$ for $b = b_\mathrm{max}$. Consequently, it spans the whole range $(-1,0)$ for $b \in (0, b_\mathrm{max})$. The $k$ bifurcation points correspond to the values of $j$ in Eq.\ (\ref{separation_theta}) equal $j = 1, \dots, k$, each yielding $\lambda = 0$. This concludes the proof of Corollary \ref{cor1}. Our numerical results obtained for $T = 5\pi$ and $T = 7\pi$ suggest that the lowest eigenvalue $\lambda_\theta$ is in fact strictly increasing with $b$; consequently, there are exactly $k$ bifurcation points on branch (b).

It is also not surprising (but of course highly nontrivial) that the number $k$ coincides with the number of periodic solutions obtained for $b = 0$ in Sec.\ \ref{bzero}. This suggests that each of the bifurcating branches of solutions ends at a separate periodic solution at $b = 0$. We do not have any strict proof of this fact, but it is confirmed by the numerical results presented in Sec. \ref{sec_bifurcation}.

Finally, note that there are no negative eigenvalues for $\phi \equiv 0$. This agrees with the fact that the branch (a) that originates at $\phi \equiv 0$ at $b = 0$ is stable.

\subsubsection{Linearization around $\phi = \phi(\alpha)$}

The case with $b = 0$ and $\phi = \phi(\alpha)$ is also relatively simple, as the potential $V(\alpha)$ is now known explicitly [here $\phi$ is given by Eq.\ (\ref{sol1})]. Separation of variables leads to a one-dimensional Schr\"{o}dinger equation. Assuming $v(\alpha,\theta) = v_1(\alpha) v_2(\theta)$ we get, from Eq.\ (\ref{schr2}),
\begin{equation}
\label{schr_one_dim}
-\partial_{\alpha \alpha} v_1 + V(\alpha) v_1 = \lambda_\alpha v_1
\end{equation}
and
\[ \frac{d}{d\mu}\left[ (1 - \mu^2) \frac{d}{d \mu} v_2 \right] + \left( \lambda - \lambda_\alpha - \frac{m^2}{1 - \mu^2} \right) v_2 = 0, \]
where again $\mu = \cos \theta$. Consequently, $\lambda = \lambda_\alpha + n(n+1)$, $n = 0, 1, \dots$

Equation (\ref{schr_one_dim}) has the form
\[ -\partial_{\alpha \alpha} v + \frac{1}{4}v - \frac{5}{4} \phi^4 v = \lambda_\alpha v \]
or
\begin{equation}
\label{eq_schr}
-\partial_{\alpha \alpha} v  - \frac{5}{4} \phi^4 v = \left( \lambda_\alpha - \frac{1}{4} \right) v,
\end{equation}
which is a one-dimensional Schr\"{o}dinger equation with the potential $\tilde V = -\frac{5}{4}\phi^4$. Surprisingly, for the limiting solution (\ref{sol2}) one gets $\tilde V = - \frac{15}{4} \mathrm{sech}^2 (\alpha - \alpha_0)$, which is a P\"{o}schl--Teller potential, for which the one-dimensional Schr\"{o}dinger equation is solvable \cite{pt}. In general, for the potential $\tilde V = - l(l+1)\mathrm{sech}^2 \alpha$, negative eigenvalues of the corresponding Schr\"{o}dinger equation are given by $-(l - j)^2$, $j = 0, \dots, N$, where $N$ is the largest integer less than $l$. For $\tilde V = - \frac{15}{4} \mathrm{sech}^2 (\alpha)$, Eq.\ (\ref{eq_schr}) has a unique negative eigenvalue $(\lambda_\alpha - 1/4) = -9/4$ or, equivalently, $\lambda_\alpha = - 2$.

\subsection{General numerical method}
\label{lin_general_method}

In the general case, we compute solutions of the eigenvalue problem Eq.\ (\ref{schr3}) numerically, using the following variant of the Riesz method. We search for the solution by expanding it in $y_{kl}$, given by Eqs.\ (\ref{ya}) and (\ref{yb}), and
\[ z_{kl} = \sin \left( k \frac{2 \pi}{T} \alpha \right) P_l (\cos \theta), \quad k = 1, \dots, M, \quad l = 0, \dots, N,  \]
i.e., we assume $v(\alpha, \theta)$ in the form
\begin{equation}
\label{expansion2}
v(\alpha, \theta) = \sum_{k=0}^M \sum_{l=0}^N a_{kl} y_{kl} (\alpha, \theta) + \sum_{k=1}^M \sum_{l=0}^N b_{kl} z_{kl} (\alpha, \theta).
\end{equation}
The functions $y_{kl}$ and $z_{kl}$ are the eigenfunctions of the operator
\[ \tilde L v \equiv - \partial_{\alpha \alpha} v - \frac{1}{\sin \theta} \partial_\theta \left( \sin \theta \partial_\theta v \right) \]
satisfying
\[ \tilde L y_{kl} = \left[ \left( \frac{2 \pi}{T} \right)^2 k^2 + l (l+1) \right] y_{kl},  \]
\[ \tilde L z_{kl} = \left[ \left( \frac{2 \pi}{T} \right)^2 k^2 + l (l+1) \right] z_{kl}. \]
Inserting expansion (\ref{expansion2}) into Eq.\ (\ref{schr}) we obtain the relation
\begin{eqnarray}
\sum_{k=0}^M \sum_{l=0}^N \left[ \left( \frac{2 \pi}{T} \right)^2 k^2 + l (l+1) + V_m(\alpha,\theta) - \lambda \right] a_{kl} y_{kl} + && \\
\sum_{k=1}^M \sum_{l=0}^N \left[ \left( \frac{2 \pi}{T} \right)^2 k^2 + l (l+1) + V_m(\alpha,\theta) - \lambda \right] b_{kl} z_{kl} & = & 0.
\end{eqnarray}
It is convenient to project the above equation on the functions
\[ \cos \left( k^\prime \frac{2 \pi}{T} \alpha \right) P_{l^\prime} (\cos \theta), \quad k^\prime = 0, \dots, M, \quad l^\prime = 0, \dots, N, \]
and
\[ \sin \left( k^\prime \frac{2 \pi}{T} \alpha \right) P_{l^\prime} (\cos \theta), \quad k^\prime = 1, \dots, M, \quad l^\prime = 0, \dots, N, \]
exploiting the fact that $V_m(\alpha, \theta)$ can be chosen as a symmetric function in $\alpha$. Then
\[ \int_0^T d \alpha \sin \left( k^\prime \frac{2 \pi}{T} \alpha \right) V_m(\alpha, \theta) y_{kl}(\alpha,\theta) = 0,  \]
and
\[ \int_0^T d \alpha \cos \left( k^\prime \frac{2 \pi}{T} \alpha \right) V_m(\alpha, \theta) z_{kl}(\alpha,\theta) = 0. \]
This leads to the following algebraic eigenvalue problems:
\begin{equation}
\label{lina}
\sum_{k=0}^M \sum_{l=0}^N \left\{ \tilde r_{k^\prime l^\prime k l} + \delta_{kk^\prime} \delta_{l l^\prime} \left[ \left( \frac{2 \pi}{T} \right)^2 k^2 + l(l+1) - \lambda \right] \right\} a_{kl} = 0,
\end{equation}
$k^\prime = 0, \dots, M$, $l^\prime = 0, \dots, N$, where
\[ \tilde r_{k^\prime l^\prime k l} = \frac{2 l + 1}{T} \int_0^\pi d \theta \sin \theta \int_0^T d \alpha P_{l^\prime}(\cos \theta) \cos\left( k^\prime \frac{2 \pi}{T} \alpha \right) V_m(\alpha,\theta) y_{kl}, \]
and
\begin{equation}
\label{linb}
\sum_{k=1}^M \sum_{l=0}^N \left\{ \tilde s_{k^\prime l^\prime k l} + \delta_{kk^\prime} \delta_{l l^\prime} \left[ \left( \frac{2 \pi}{T} \right)^2 k^2 + l(l+1) - \lambda \right] \right\} b_{kl} = 0,
\end{equation}
$k^\prime = 1, \dots, M$, $l^\prime = 0, \dots, N$, where
\[ \tilde s_{k^\prime l^\prime k l} = \frac{2 l + 1}{T} \int_0^\pi d \theta \sin \theta \int_0^T d \alpha P_{l^\prime}(\cos \theta) \sin\left( k^\prime \frac{2 \pi}{T} \alpha \right) V_m(\alpha,\theta) y_{kl}. \]
Both eigenvalue equations (\ref{lina}) and (\ref{linb}) can be solved easily. In practice, it is convenient to introduce `one dimensional' indices numbering the expansion functions. This allows one to rewrite Eqs.\ (\ref{lina}) and (\ref{linb}) in the standard matrix notation. For the indices $k = 0, \dots, M$ and $l = 0, \dots, N$ we introduce the index $J = k(N+1) + l + 1$, with the inverse relation given by
\[ k = \lfloor \frac{J - 1}{N + 1} \rfloor, \quad l = J - 1 - k(N+1), \]
where $\lfloor x \rfloor$ denotes the largest integer less than or equal to $x$. For the indices $k = 1, \dots, M$ and $l = 0, \dots, N$ we define $J = (k - 1)(N + 1) + l + 1$, with the inverse
\[ k = \lfloor \frac{J - 1}{N + 1} \rfloor + 1, \quad l = J - 1 - (k - 1)(N+1). \]

\subsection{Numerical results}
\label{lin_results}

\begin{figure}[t]
\begin{center}
\includegraphics[width=0.8\textwidth]{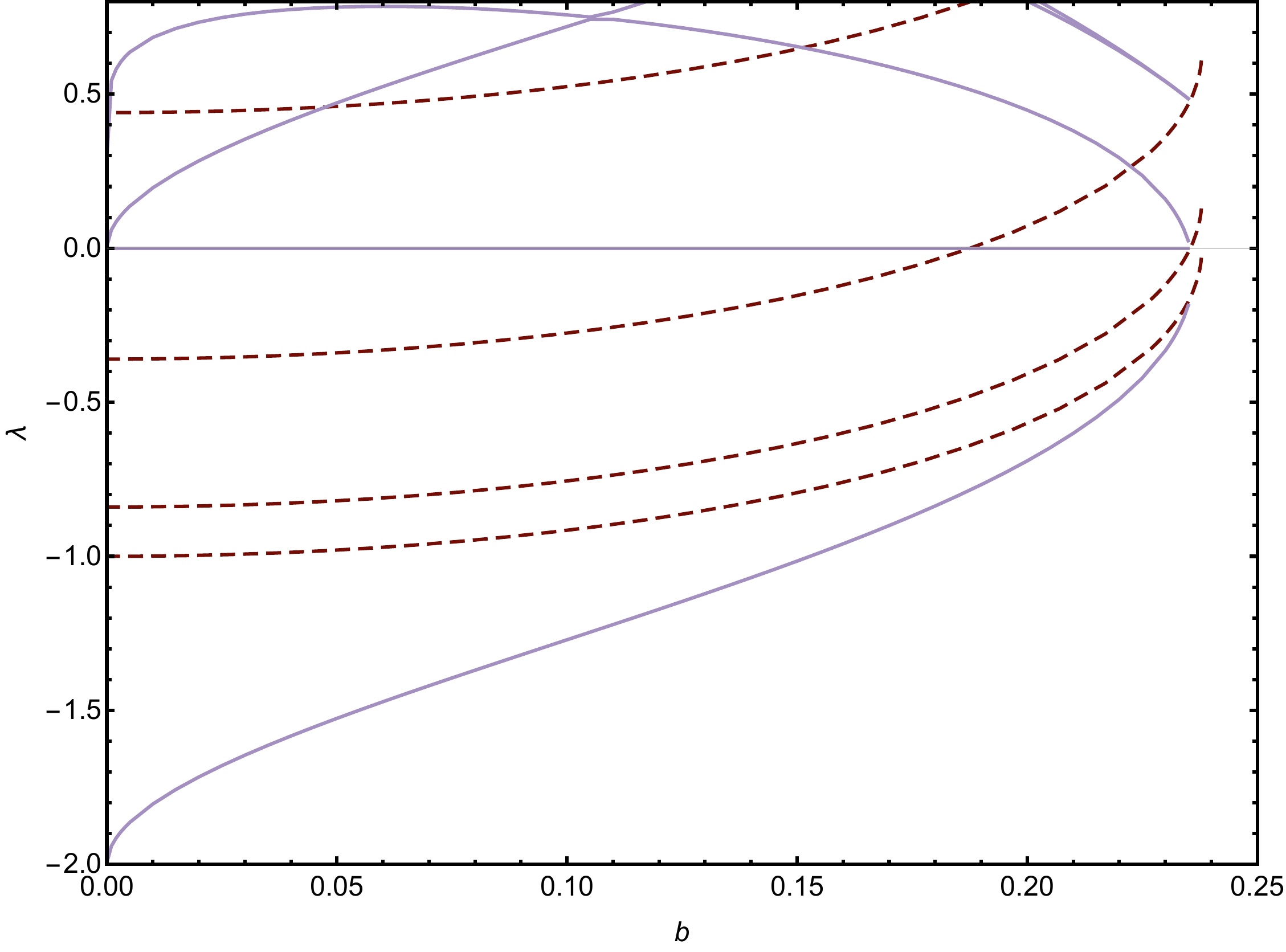}
\end{center}
\caption{\label{widmoa}Lowest eigenvalues of the linearized Eq.\ (\ref{schr2}) corresponding to branches (b) and (c). Here $T = 5\pi$, $m = 0$.}
\end{figure}

\begin{figure}[t]
\begin{center}
\includegraphics[width=0.8\textwidth]{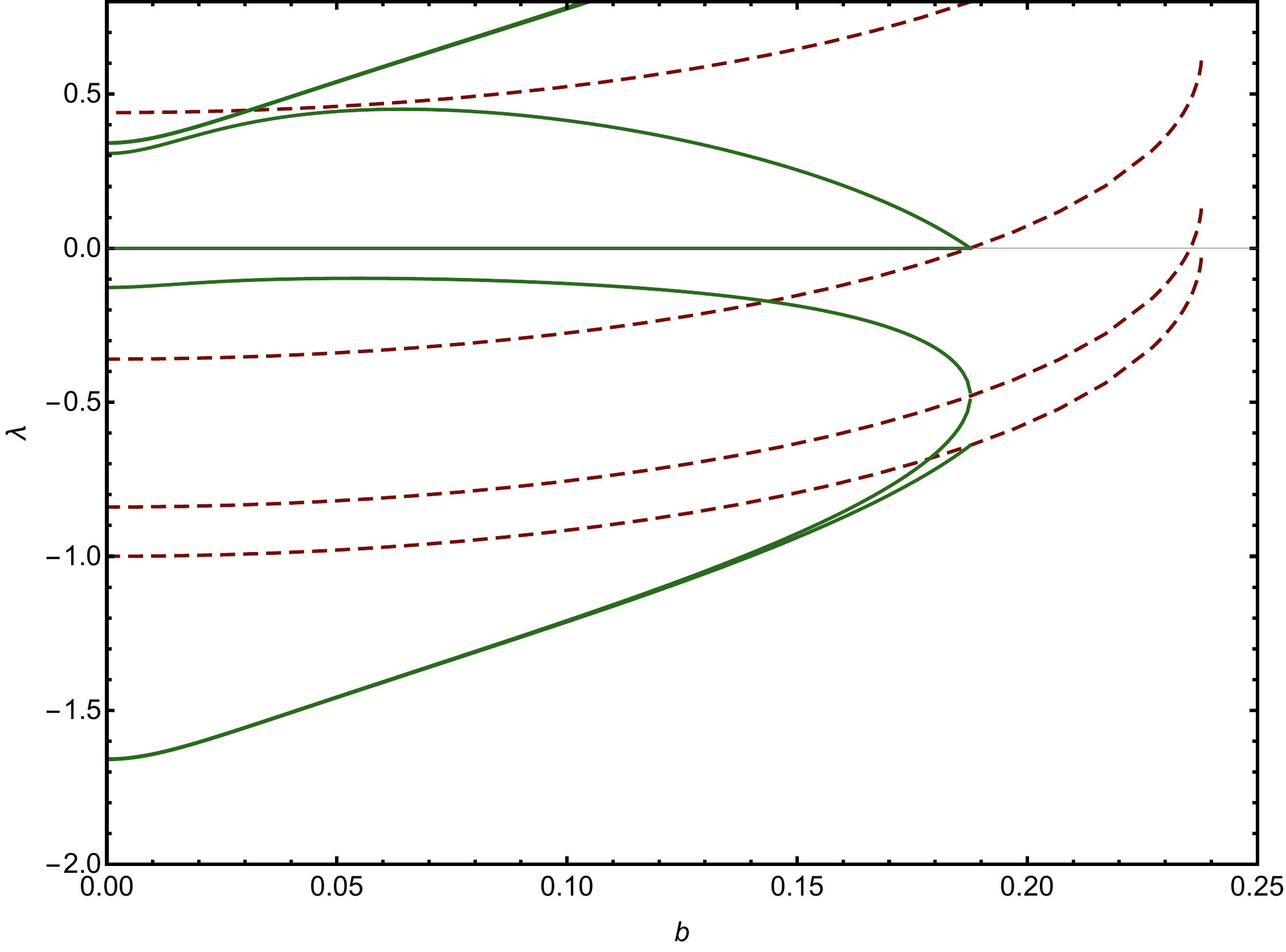}
\end{center}
\caption{\label{widmob}Same as in Fig.\ \ref{widmoa}, but for branches (b) and (d).}
\end{figure}

\begin{figure}[t]
\begin{center}
\includegraphics[width=0.8\textwidth]{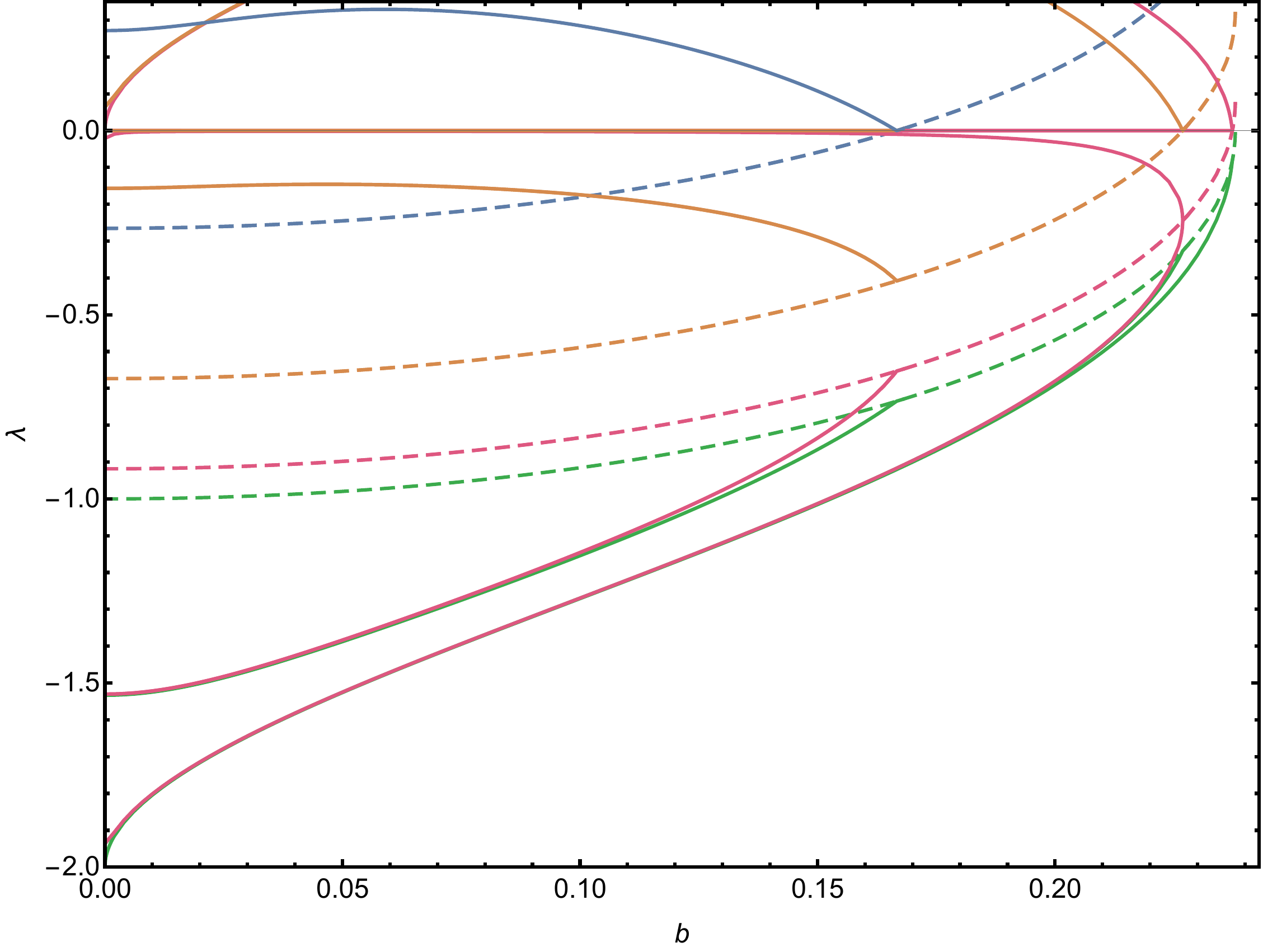}
\end{center}
\caption{\label{widmoc}Lowest eigenvalues of the linearized Eq.\ (\ref{schr2}). Here $T = 7 \pi$, $m = 0$.}
\end{figure}

\begin{figure}[t]
\begin{center}
\includegraphics[width=0.8\textwidth]{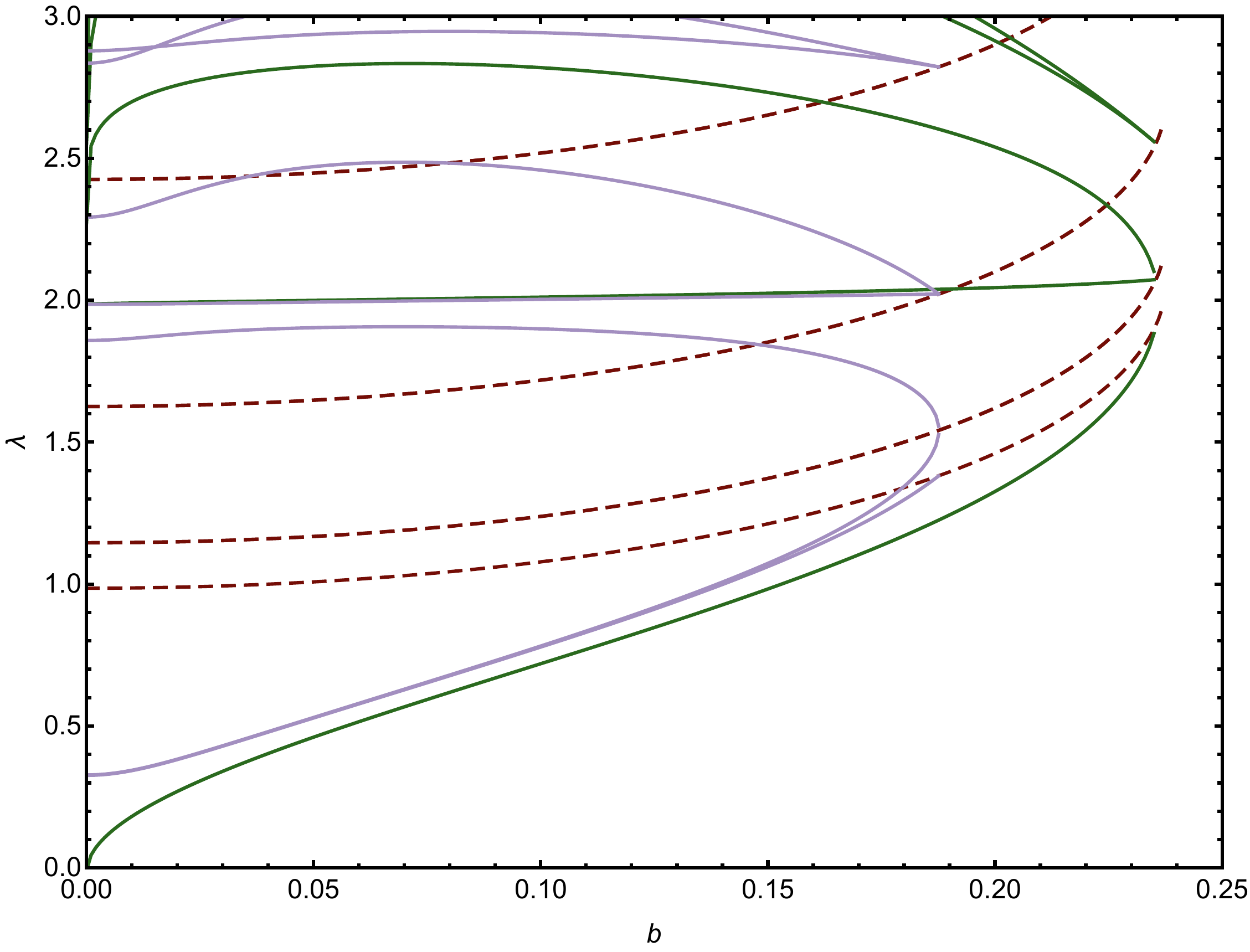}
\end{center}
\caption{\label{m1}Lowest eigenvalues of the linearized Eq.\ (\ref{schr2}). Here $T = 5 \pi$, $m = 1$.}
\end{figure}

We will now discuss numerical solutions of the linearized equations introduced in Sec.\ \ref{linear}. The purpose of this analysis is twofold. Firstly, we would like to illustrate the origin of the already known bifurcations in the system --- they should correspond to the occurrence of zero modes of the linearized equation. Secondly, we would like to collect numerical evidence suggesting that no additional bifurcation points exist. As before we restrict ourselves to sample periods $T = 5 \pi$ and $T = 7 \pi$, and solutions of the Lichnerowicz equation that were already discussed in Sec.\ \ref{sec_sols}.

In practice, we solve Eqs.\ (\ref{schr2}) or (\ref{schr3}) with the potential $V$ determined by a given numerical solution $\phi$. Figures \ref{widmoa} and \ref{widmob} show examples of the eigenvalues corresponding to branches (b), (c) and (d), discussed in Sec.\ \ref{sec_bifurcation}. We assumed $T = 5 \pi$ and $m = 0$. In both graphs the abscissa corresponds to the parameter $b$; the ordinate gives appropriate eigenvalues. Since we are interested in the occurrence of zero modes, we only plotted eigenvalues satisfying $\lambda \lessapprox 0.8$. We deliberately omit plotting strictly positive eigenvalues corresponding to the stable branch (a).

In both Figs.\ \ref{widmoa} and \ref{widmob} we plot eigenvalues corresponding to branch (b). They are denoted with dotted lines. Solid lines in Figs.\ \ref{widmoa} and \ref{widmob} denote, respectively, eigenvalues corresponding to bifurcating branches (c) and (d). 

Branch (b) originates at $b = 0$ with a limiting solution $\phi \equiv 1$, and ends at the maximum value of the parameter $b = b_\mathrm{max}$. It consists of solutions $\phi$ that depend only on $\theta$. Consequently, one expects to find a spectrum predicted by Eq.\ (\ref{separation_theta}). The branch of lowest eigenvalues $\lambda_0 = \lambda_\theta$ starts at $b = 0$ with the value $\lambda_0 = \lambda_\theta = -1$, as given by Eq.\ (\ref{eigen_phi_1}) with $j = n = 0$, and increases monotonically to $\lambda_0 = 0$ at $b = b_\mathrm{max}$. For $T = 5\pi$ higher eigenvalues are given by $\lambda_1 = \lambda_\theta + 4/25$, $\lambda_2 = \lambda_\theta + 16/25$, $\lambda_3 = \lambda_\theta + 36/25$, etc. The branch $\lambda_3$ is already strictly positive, but branches $\lambda_1$ and $\lambda_2$ have zeros at $b = b_1$ and $b = b_2$ respectively. These zeros give rise to two bifurcating branches of solutions: (c) and (d). Note that $\lambda_0 = -4/25$ at $b = b_1$. At $b = b_2$ we have $\lambda_0 = -16/25$ and $\lambda_1 = -12/25$.

Eigenvalues corresponding to branch (c) are shown in Fig.\ \ref{widmoa}. For each solution $\phi$ belonging to this branch there is a nontrivial eigenfunction $\partial_\alpha \phi$ and an eigenvalue $\lambda = 0$. The only negative eigenvalue reads $\lambda_0 = -1.994$ at $b = 0$ and grows monotonically up to $\lambda_0 = -4/25$ at $b = b_1$ [it bifurcates from the $\lambda_0$ eigenvalue that corresponds to branch (b)]. The remaining eigenvalues are already nonnegative with possible zeros at $b = 0$ or $b = b_1$. Consequently they do not give rise to any bifurcating branches of solutions $\phi$.

The spectrum corresponding to branch (d) is more complex. There are 3 negative eigenvalues $\lambda_0$, $\lambda_1$, and $\lambda_2$. The lowest eigenvalue $\lambda_0$ reads $\lambda_0 = -1.659$ at $b = 0$. It grows monotonically up to $b = b_2$, where $\lambda_0 = - 16/25$ [it bifurcates from the $\lambda_0$ eigenvalue that corresponds to branch (b)]. The two eigenvalues $\lambda_1$ and $\lambda_2$ bifurcate from the eigenvalue $\lambda_1$ corresponding to branch (b) at $b = b_2$. We have, at $b = b_2$, $\lambda_1 = \lambda_2 = - 12/25$. At $b = 0$, $\lambda_1$ and $\lambda_2$ read $\lambda_1 = -1.658$ and $\lambda_2 = -0.127$. Note that at $b = 0$ the eigenvalues $\lambda_0$ and $\lambda_1$ are almost degenerate. This is a characteristic feature of solutions of the Schr\"{o}dinger equation with a double-minimum potential\footnote{It is a well-known effect, observed already in 1927 by Hund \cite{hund}; see also the discussion and references in \cite{hodgson}.}. The eigenvalue $\lambda_1$ is monotonically growing with $b$, but $\lambda_2$ is not monotonic. Similarly to the spectrum of branch (c), there is also a zero eigenvalue corresponding to the eigenfunction $\partial_\alpha \phi$. All other eigenvalues are nonnegative.

We believe that this picture is generic. We give another example in Fig.\ \ref{widmoc}, which shows spectra analogous to those shown in Figs.\ \ref{widmoa} and \ref{widmob}, but obtained for the case with $T = 7 \pi$.

In principle, one can also search for possible zero eigenvalues for $m \neq 0$. Figure \ref{m1} shows a sample plot of the eigenvalues obtained for $m = 1$ and $T = 5\pi$. Except at $b = 0$, all eigenvalues are strictly positive; consequently, there is no need to investigate the spectra corresponding to higher values of $m$. For $m = 1$ and $b = 0$ the lowest eigenvalue $\lambda_0 = 0$. There is a numerical subtlety connected with this result. Our numerical computations actually yield a slightly negative eigenvalue $\lambda_0$ that tends to zero with an increase of the numerical accuracy. All these results support the conjecture that no additional branches of solutions depending on $\varphi$ bifurcate from branches (b), (c) and (d).

\section{Concluding remarks}
\label{conclusions}

The so-called conformal method is probably the best known way of solving the Einstein constraint equations. In some form it was present already in the work of Lichnerowicz \cite{lichnerowicz}. A more recent version of the conformal method can be found in \cite{york, niall_york}. We believe that the investigation of the Bowen-York initial data can provide new insights into the nature of the initial value problem in the case with a positive cosmological constant.

Here, we only dealt with the `rotating' case, for which Bizo\'{n}, Pletka and Simon introduced a particularly elegant and simple generalization. In this approach the resulting Lichnerowicz equation depends explicitly only on one variable --- the polar coordinate $\theta$. The compactification of the `radial' variable $\alpha$ is another restriction. It is a remarkable property that there exist solutions periodic with respect to $\alpha$, which, in a sense, justify calling such a compactification a `natural' one. In any case, the properties and the classification of solutions depend strongly on the fact that we are dealing with a Lichnerowicz equation on a compact domain.

Except for the case with $b = 0$ and the proof of the existence of symmetry-breaking bifurcations, the results presented in this work are mostly numerical. We focused on new branches of solutions, whose existence was predicted by Bizo\'{n}, Pletka and Simon in \cite{BPS}. It should be also noted that proving the completeness of the pattern of solutions presented in this paper remains an open problem. We hope that the numerical evidence given in the second part of this paper can motivate further research in this direction.

\section*{Acknowledgement}

We would like to thank Piotr Bizo\'{n} and Walter Simon for many stimulating discussions concerning this project. PM acknowledges the financial support of the Narodowe Centrum Nauki Grant No.\ DEC-2012/06/A/ST2/00397. JK was supported by the Polish Ministry of Science and Higher Education within the `Diamentowy Grant' program (No. 0153/DIA/2016/45).

\section*{Appendix: Gauss--Legendre--Fourier quadratures}

Gauss--Legendre and Gauss--Fourier quadratures are frequently used in the implementation of spectral methods. Here we recall appropriate formulae for completeness.

Let us define $x = \cos \theta$. The task is to compute integrals of the form
\begin{eqnarray*}
I_{kl} & \equiv & \frac{2l+1}{T} \int_0^\pi d\theta \sin \theta \int_0^T d\alpha f(\alpha, \theta) \cos\left( k \frac{2\pi}{T} \alpha \right) P_l(\cos \theta) \\
& = & \frac{2l+1}{T} \int_{-1}^1 dx \int_0^T d\alpha f(\alpha, x) \cos\left( k \frac{2\pi}{T} \alpha \right) P_l(x),
\end{eqnarray*}
where for simplicity no separate symbol is reserved for the function $f$ in the new coordinate system. Integration with respect to $x$ can be performed by introducing the $N$ zeros of the polynomial $P_N$ (they are denoted by $x_i$, $i = 1, \dots, N$) and $N$ weights
\[ w_i = \frac{2(1 - x_i^2)}{(N+1)^2 P_{N+1}^2(x_i)}, \quad i = 1, \dots, N. \]
Integration with respect to $\alpha$ requires $\tilde M = 2M + 1$ collocation points $\alpha_j = T (j - 1)/\tilde M$, $j = 1, \dots, \tilde M$. The integral $I_{kl}$ is then computed as
\[ I_{kl} = \frac{2l + 1}{\tilde M} \sum_{j = 1}^{\tilde M} \sum_{i = 1}^N f(\alpha_j,x_i) \cos\left( k \frac{2\pi}{T} \alpha_j \right) w_i P_l(x_i). \]
The above formula is exact for all functions $f$ that can be spanned by $y_{kl}$, $k = 0, \dots, M$, $l = 0, \dots, N$.
Similarly, the integral
\begin{eqnarray*}
J_{kl} & \equiv & \frac{2l+1}{T} \int_0^\pi d\theta \sin \theta \int_0^T d\alpha f(\alpha, \theta) \sin\left( k \frac{2\pi}{T} \alpha \right) P_l(\cos \theta) \\
& = & \frac{2l+1}{T} \int_{-1}^1 dx \int_0^T d\alpha f(\alpha, x) \sin\left( k \frac{2\pi}{T} \alpha \right) P_l(x),
\end{eqnarray*}
is computed as
\[ J_{kl} = \frac{2l + 1}{\tilde M} \sum_{j = 1}^{\tilde M} \sum_{i = 1}^N f(\alpha_j,x_i) \sin\left( k \frac{2\pi}{T} \alpha_j \right) w_i P_l(x_i). \]

\end{document}